\documentclass[longauth]{aa}
\usepackage[utf8]{inputenc}
\usepackage[varg]{txfonts}
\usepackage[usenames, dvipsnames]{color}
\usepackage{amsmath}
\usepackage{graphicx}
\usepackage[english]{babel}
\usepackage{siunitx}
\usepackage{float}
\usepackage{url}
\usepackage{longtable}
\usepackage{fancyhdr}
\usepackage{natbib}
\usepackage[normalem]{ulem}

\newcommand{\Tzerob}[1][days]{$8180.2842 _{ - 0.0015 } ^ { + 0.0017 } $#1} 
\newcommand{\Pb}[1][days]{$16.9841 _{ - 0.0008 } ^ { + 0.0008 } $#1} 
\newcommand{\eb}[1][ ]{$0.04 _{ - 0.03 } ^ { + 0.06 } $#1} 
\newcommand{\wb}[1][deg]{$104 _{ - 35 } ^ { + 22 } $#1} 
\newcommand{\bb}[1][ ]{$0.68 _{ - 0.08 } ^ { + 0.06 } $#1} 
\newcommand{\arb}[1][ ]{$25.9 _{ - 2.0 } ^ { + 1.8 } $#1} 
\newcommand{\rrb}[1][ ]{$0.0220 _{ - 0.0005 } ^ { + 0.0006 } $#1} 
\newcommand{\kb}[1][${\rm m\,s^{-1}}$]{$6.1 _{ - 1.1 } ^ { + 1.1 } $#1} 
\newcommand{\mpb}[1][$M_{\oplus}$]{$24.5 _{ - 4.4 } ^ { + 4.4 } $#1} 
\newcommand{\rpb}[1][$R_{\oplus}$]{$2.63 _{ - 0.10 } ^ { + 0.11 } $#1} 
 
\newcommand{\ib}[1][deg]{$88.4 _{ - 0.3 } ^ { + 0.2 } $#1} 
\newcommand{\ab}[1][au]{$0.13 _{ - 0.01 } ^ { + 0.01 } $#1} 
\newcommand{\insolationb}[1][${\rm F_{\oplus}}$]{$67.0 _{ - 9.0 } ^ { + 12.0 } $#1} 
\newcommand{\denstrb}[1][${\rm g\,cm^{-3}}$]{$1.14 _{ - 0.24 } ^ { + 0.26 } $#1} 
 
\newcommand{\denpb}[1][${\rm g\,cm^{-3}}$]{$7.4 _{ - 1.5 } ^ { + 1.6 } $#1} 
 
\newcommand{\grapparsb}[1][${\rm cm\,s^{-2}}$]{$3460.0 _{ - 660.0 } ^ { + 680.0 } $#1} 
 
\newcommand{\Teqb}[1][K]{$795 _{ - 28 } ^ { + 33 } $#1} 
\newcommand{\ttotb}[1][hours]{$3.66 _{ - 0.08 } ^ { + 0.07 } $#1} 
 
\newcommand{\qone}[1][]{$0.36 _{ - 0.16 } ^ { + 0.25 } $#1} 
\newcommand{\qtwo}[1][]{$0.26 _{ - 0.20 } ^ { + 0.35 } $#1} 
\newcommand{\uone}[1][]{$0.31 _{ - 0.23 } ^ { + 0.31 } $#1} 
\newcommand{\utwo}[1][]{$0.27 _{ - 0.27 } ^ { + 0.35 } $#1} 
\newcommand{\CARMENES}[1][${\rm m\,s^{-1}}$]{$48 _{ - 8 } ^ { + 8 } $#1} 
\newcommand{\ltrend}[1][${\rm m\,s^{-1}\,d^{-1}}$]{$-0.40 _{ - 0.07 } ^ { + 0.07 } $#1}

\title{Detection and characterization of an ultra-dense sub-Neptune planet orbiting the Sun-like star HD~119130}
\subtitle{}
\titlerunning{An ultra-dense sub-Neptune planet orbiting the Sun-like star HD~119130}

\author{R.~Luque\inst{1,2}
   \and G.~Nowak\inst{1,2}
   \and E.~Pallé \inst{1,2}
   \and F.~Dai\inst{3,4}%
   \and A.~Kaminski\inst{5}
   \and E.~Nagel\inst{6}
   \and D.~Hidalgo\inst{1,2}%
   \and F.~Bauer\inst{7}
   \and M.~Lafarga\inst{8,9} 
   \and J.~Livingston\inst{10}%
   \and O.~Barrag\'an\inst{11}%
   \and T.~Hirano\inst{12}%
   \and M.~Fridlund\inst{13,14}%
   \and D.~Gandolfi\inst{12}%
   \and A.\,B.~Justesen\inst{15}%
   \and M.~Hjorth \inst{15}%
   \and V.~Van~Eylen\inst{3}%
   \and J.\,N.~Winn\inst{3}%
   \and M.~Esposito\inst{16}%
   \and J.\,C.~Morales\inst{8,9}%
  \and S.~Albrecht\inst{15}%
  \and R.~Alonso\inst{1,2}%
  \and P.\,J.~Amado\inst{7}
  \and P.~Beck\inst{1,2}
  \and J.~A.~Caballero\inst{17}
  \and J.~Cabrera\inst{18}%
  \and W.\,D.~Cochran\inst{19}%
  \and Sz.~Csizmadia\inst{18}%
  \and H.~Deeg\inst{1,2}%
  \and Ph.~Eigm\"uller\inst{18}%
  \and M.~Endl\inst{19}%
  \and A.~Erikson\inst{18}%
  \and A.~Fukui\inst{20}%
  \and S.~Grziwa\inst{21}%
  \and E.\,W.~Guenther\inst{16}%
  \and A.\,P.~Hatzes \inst{16}%
  \and E.~Knudstrup\inst{15}
  \and J.~Korth\inst{21}%
  \and K.\,W.\,F.~Lam\inst{22}%
  \and M.\,N.~Lund\inst{15}%
  \and S.~Mathur\inst{1,2}%
  \and P.~Monta\~nes-Rodr\'iguez\inst{1,2}%
  \and N.~Narita\inst{23,11,1}%
  \and D.~Nespral \inst{1,2}%
  \and P.~Niraula \inst{24}%
  \and M.~P\"atzold\inst{21}%
  \and C.\,M.~Persson \inst{13}%
  \and J.~Prieto-Arranz\inst{1,2}%
  \and A.~Quirrenbach\inst{5}%
  \and H.~Rauer\inst{18,22,25}%
  \and S.~Redfield\inst{26}%
  \and A.~Reiners\inst{27}
  \and I.~Ribas\inst{8,9}%
  \and A.\,M.\,S.~Smith\inst{18}%
}

\institute{
Instituto de Astrof\'isica de Canarias (IAC), 38205 La Laguna, Tenerife, Spain;
\email{rluque@iac.es}
\and 
Departamento de Astrof\'isica, Universidad de La Laguna (ULL), 38206, La Laguna, Tenerife, Spain 
\and
Department of Astrophysical Sciences, Princeton University, 4 Ivy Lane, Princeton, NJ 08544, USA 
\and 
Department of Physics and Kavli Institute for Astrophysics and Space Research, Massachusetts Institute of Technology, Cambridge, MA 02139, USA 
\and 
Landessternwarte, Zentrum für Astronomie der Universtät Heidelberg, Königstuhl 12, D-69117 Heidelberg, Germany
\and 
Hamburger Sternwarte, Gojenbergsweg 112, D-21029 Hamburg, Germany
\and 
Instituto de Astrofísica de Andalucía (IAA-CSIC), Glorieta de la Astronomía s/n, E-18008 Granada, Spain
\and
Institut de Ci\`encies de l\'~Espai (ICE, CSIC), C/Can Magrans, s/n, Campus UAB, 08193 Bellaterra, Spain 
\and
Institut d’Estudis Espacials de Catalunya (IEEC), E-08034 Barcelona, Spain 
\and
Department of Astronomy, The University of Tokyo, 7-3-1 Hongo, Bunkyo-ku, Tokyo 113-0033, Japan 
\and
Dipartimento di Fisica, Universit\`a di Torino, Via P. Giuria 1, I-10125, Torino, Italy 
\and
Department of Earth and Planetary Sciences, Tokyo Institute of Technology, 2-12-1 Ookayama, Meguro-ku, Tokyo 152-8551, Japan 
\and
Department of Space, Earth and Environment, Chalmers University of Technology, Onsala Space Observatory, 439 92 Onsala, Sweden 
\and
Leiden Observatory, Leiden University, 2333CA Leiden, The Netherlands 
\and 
Stellar Astrophysics Centre, Department of Physics and Astronomy, Aarhus University, Ny Munkegade 120, DK-8000 Aarhus C, Denmark 
\and 
Th\"uringer Landessternwarte Tautenburg, Sternwarte 5, 07778 Tautenburg, Germany 
\and 
Centro de Astrobiología (CSIC-INTA), ESAC campus, Camino Bajo del Castillo s/n, E-28692 Villanueva de la Cañada, Madrid, Spain
\and
Institute of Planetary Research, German Aerospace Center, Rutherfordstrasse 2, 12489 Berlin, Germany 
\and 
Department of Astronomy and McDonald Observatory, University of Texas at Austin, 2515 Speedway, Stop C1400, Austin, TX 78712, USA 
\and 
Okayama Astrophysical Observatory, National Astronomical Observatory of Japan, NINS, Asakuchi, Okayama 719-0232, Japan 
\and 
Rheinisches Institut f\"ur Umweltforschung an der Universit\"at zu K\"oln, Aachener Strasse 209, 50931 K\"oln, Germany 
\and 
Zentrum f\"ur Astronomie und Astrophysik, Technische Universit\"at Berlin, Hardenbergstr. 36, 10623 Berlin, Germany
\and 
Astrobiology Center and National Astronomical Observatory of Japan, NINS, 2-21-1 Osawa, Mitaka, Tokyo 181-8588, Japan 
\and
Department of Earth, Atmospheric and Planetary Sciences, MIT, 77 Massachusetts Avenue, Cambridge, MA 02139
\and
Institute of Geological Sciences, Freie Universit\"at Berlin, Malteserstr. 74-100, 12249 Berlin, Germany
\and
Astronomy Department and Van Vleck Observatory, Wesleyan University, Middletown, CT 06459, USA 
\and
Institut für Astrophysik, Georg-August-Universität, Friedrich-Hund-Platz 1, D-37077 Göttingen, Germany
}

\date{Received 21 Dec 2018 / Accepted dd MM 2019}

\abstract{We present the discovery and characterization of a new transiting planet from Campaign 17 of the {\it Kepler} extended mission {\it K2}. HD~119130~b is a warm sub-Neptune on a 17-d orbit around a bright ($V=9.9\,\mathrm{mag}$) solar-like G3~V star with a mass and radius of $M_\star = 1.00\pm0.03\,\mathrm{M_\odot}$ and $R_\star = 1.09\pm0.03\,\mathrm{R_\odot}$, respectively. We model simultaneously the {\it K2} photometry and CARMENES spectroscopic data and derive a radius of $R_\mathrm{p} = 2.63_{-0.10}^{+0.12}\,\mathrm{R_\oplus}$ and mass of $M_\mathrm{p} = 24.5_{-4.4}^{+4.4}\,\mathrm{M_\oplus}$, yielding a mean density of $\rho_\mathrm{p} = 7.4_{-1.5}^{+1.6}\,\mathrm{g\,cm^{-3}}$, which makes it one of the densest sub-Neptune planets known to date. We also detect a linear trend in radial velocities of HD~119130 ($\dot{\gamma}_{\rm RV}= -0.40^{+0.07}_{-0.07}\,\mathrm{m\,s^{-1}\,d^{-1}}$) that suggests a long-period companion with a minimum mass on the order of $33\,\mathrm{M_\oplus}$. If confirmed, it would support a formation scenario of HD~119130~b by migration caused by Kozai-Lidov oscillations.}


\begin{document}

\maketitle

\section{Introduction}
The synergy between the first ground-based transit surveys and high precision velocimeters aimed at detecting bona fide planets, has been a first step towards comprehensive characterizations of exoplanets. The extended {\it Kepler} mission {\it K2} \citep{2014PASP..126..398H} was the first to provide the opportunity to search for transiting planets with radii smaller than Neptune and larger than Earth around a significant number of bright, solar-type stars amenable to high precision radial velocity (RV) measurements. This opened the door to the precise determination of masses and, hence, bulk densities of this new class of planets that have no counterparts in the Solar System.

Results from the original {\it Kepler} mission \citep{2010Sci...327..977B} showed that planets with radii smaller than Neptune are the most frequently occurring type within 1\,au of solar-type stars \citep[e.g.][]{Petigura18}. Besides, a quarter of all Sun-like stars --up to 50\% in the recent study of \citet{Petigura18}-- hosts planets smaller than $4\,\mathrm{R_\oplus}$ with orbital periods shorter than 100 days \citep{2013ApJS..204...24B,2014PNAS..11112655M,2015ApJ...799..180S}. Their radius distribution shows a bi-modal structure with a gap around $1.7\,\mathrm{R_\oplus}$ that separates presumably rocky super-Earths, with radii centered at $1.2\,\mathrm{R_\oplus}$, from gas-dominated sub-Neptunes, with radii centered at $2.4\,\mathrm{R_\oplus}$ \citep{Fulton17,Fulton18}. The gap has also a potential slope as a function of orbital period and a possible decrease in the value of the critical radius with increasing period \citep{2018MNRAS.479.4786V}.

While the shape of the bi-modal radius distribution can be understood with photoevaporation models \citep{2013ApJ...775..105O,2013ApJ...776....2L,2017ApJ...847...29O}, it is still not clear how close-in planets are formed. Different mechanisms have been studied, and they can be roughly divided in two categories: in-situ formation and migration. \citet{2012ApJ...751..158H}, among others, suggested that low-mass super-Earths/sub-Neptunes formed close to their current orbital locations. On the other hand, disk-driven migration could also explain the formation of close-in planets, e.g. \citet{2009A&A...501.1139M,2009A&A...501.1161M,2014Sci...346..212D}. However, little is known for the rare population of ultra-dense sub-Neptunes without significant envelopes.

In this paper, we present the detection and thorough characterization in terms of radius, mass, and mean density of one such planet, along with a characterization of its host star, HD~119130. Analysis of the {\it K2} photometry presented in Section~\ref{sec:k2} reveals that HD~119130~b belongs to the group of presumably gas-dominated sub-Neptunes. In Section~\ref{sec:obs} we present high-resolution images that allow us to exclude a false-positive scenario due to the background/foreground or physically bound eclipsing binaries. In addition, we present CARMENES high-resolution spectroscopy, used in Section~\ref{sec:star} for precise determination of the host star's parameters that reveals its solar-like nature. A joint analysis of CARMENES high precision RVs and {\it K2} photometry performed to determine the planetary parameters is presented in Section~\ref{sec:fit}. Finally, in Section~\ref{sec:discussion}, we discuss the implications of the high density of HD~119130~b for planet composition models and formation scenarios.

\section{Candidate detection from {\it K2} photometry} \label{sec:k2}

\begin{figure}
\centering
\includegraphics[width=\hsize]{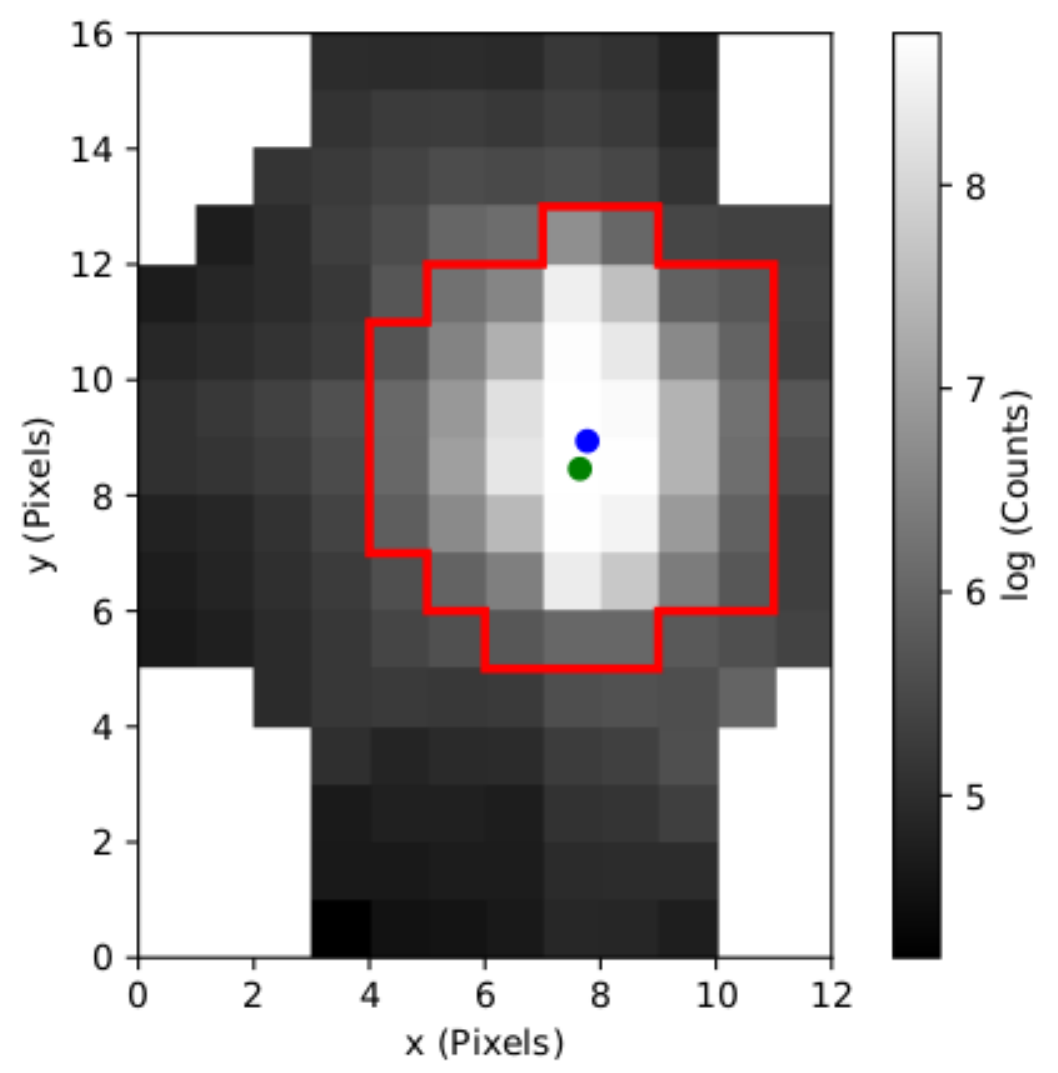}
\caption{Photometric aperture (red contour) used to construct the {\it K2} light curve. The center of the flux distribution is depicted with a blue dot, while the position of the target in the EPIC catalog is marked in green. } 
\label{fig:aperture}
\end{figure}

The {\it K2} Field 17 was observed from March 1st until May 8th 2018, in the direction of the constellation of Virgo. Among the 34\,398 long-cadence (29.4-min integration time) targets was HD~119130 (EPIC~212628254), proposed by several Guest Observer (GO) programs: GO-17003 (Cochran), GO-17032 (Buzasi), GO-17049 (Howard), and GO-17065 (Dressing). HD~119130 is a poorly-investigated star with few determined parameters (spectral type by \citealt{1999MSS...C05....0H}; RAVE kinematics and metallicities by \citealt{Boeche11,Coskunoglu12}; basic stellar parameters by \citealt{2014AJ....148...81M}). We downloaded the target pixel files from the Mikulski Archive for Space Telescopes (MAST) and extracted the light curve using two methods. 

Our first method \citep[for details see][]{Dai17} attempts to reduce the brightness fluctuations associated with the rolling motion of the spacecraft using the observed motion of the center-of-light on the detector, similar to the one described by \citet{Vanderburg14}. For each image, we set a circular aperture around the brightest pixel and fit a two-dimension Gaussian function to the flux distribution. In the case of HD~119130, the aperture with the lowest levels of noise has a radius of $\sim$\,4\,$\mathrm{pix}$, as seen in Fig.~\ref{fig:aperture}. Then, a piecewise linear function between the observed flux variation and the centroid coordinates of the Gaussian is fitted, which is used to detrend the observed intensity fluctuations. Secondly, to check the consistency in our results, we also build the light curve using an independent method based on the implementation of the pixel level decorrelation model \citep{Deming15} using a customized version of the \texttt{Everest} pipeline \citep{Luger17}. The details of this procedure are described in \citet{Palle18}. 

\begin{figure}
\centering
\includegraphics[width=\hsize]{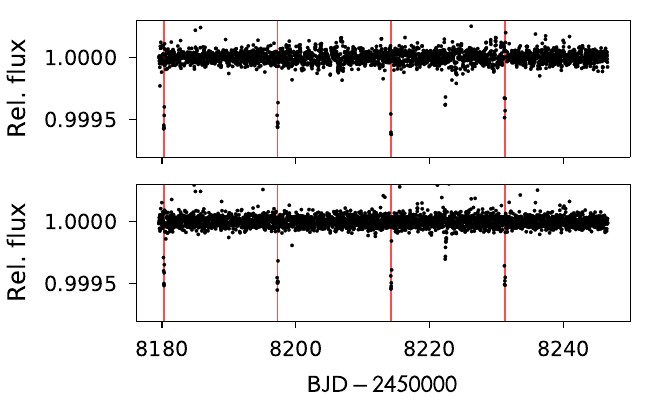}
\caption{Detrended {\it K2} light curves of HD~119130 using the two methods described in Section~\ref{sec:k2}. {\it Top panel:} \citet{Dai17} method à la \citet{Vanderburg14}. {\it Bottom panel:} \citet{Palle18} method based on the Everest pipeline \citep{Luger17}. The vertical red lines mark the position of each detected transit.} 
\label{fig:k2_lc}
\end{figure}
We searched for transits in both light curves using the Box-fitting Least Squares algorithm \citep[BLS;][]{BLS}, improved to account for the expected scaling of transit duration with orbital period as pointed by \citet{Ofir14}. Once a planetary signal is detected in the power spectrum, we remove such transit feature and reapply the BLS algorithm iteratively until no further signals are detected. The detrended light curve and transits detections by each method are shown in Fig.~\ref{fig:k2_lc}. Four transits with a depth of $\sim$\,550\,$\mathrm{ppm}$ are clearly detected, whose linear best-fit ephemeris analysis returns a planet candidate with a period of $P \sim$\,17\,$\mathrm{d}$. The flux decrement seen at $\mathrm{BJD}-2\,450\,000 \approx 8\,222$ is related with {\it K2} systematics and not to a single transit event.

In order to check for the possibility of a binary scenario that is mimicking the observed transit signal we searched the light curves for odd-even transit depth variations and secondary eclipse features. The depth of the odd-even transits agrees within $1\sigma$ and there is no hint of a secondary eclipse. As a result, we triggered a follow-up campaign to characterize precisely the HD~119130 system.

\section{Ground-based follow-up observations} \label{sec:obs}

\subsection{High-resolution imaging}
\begin{figure}
\centering
\includegraphics[width=\hsize]{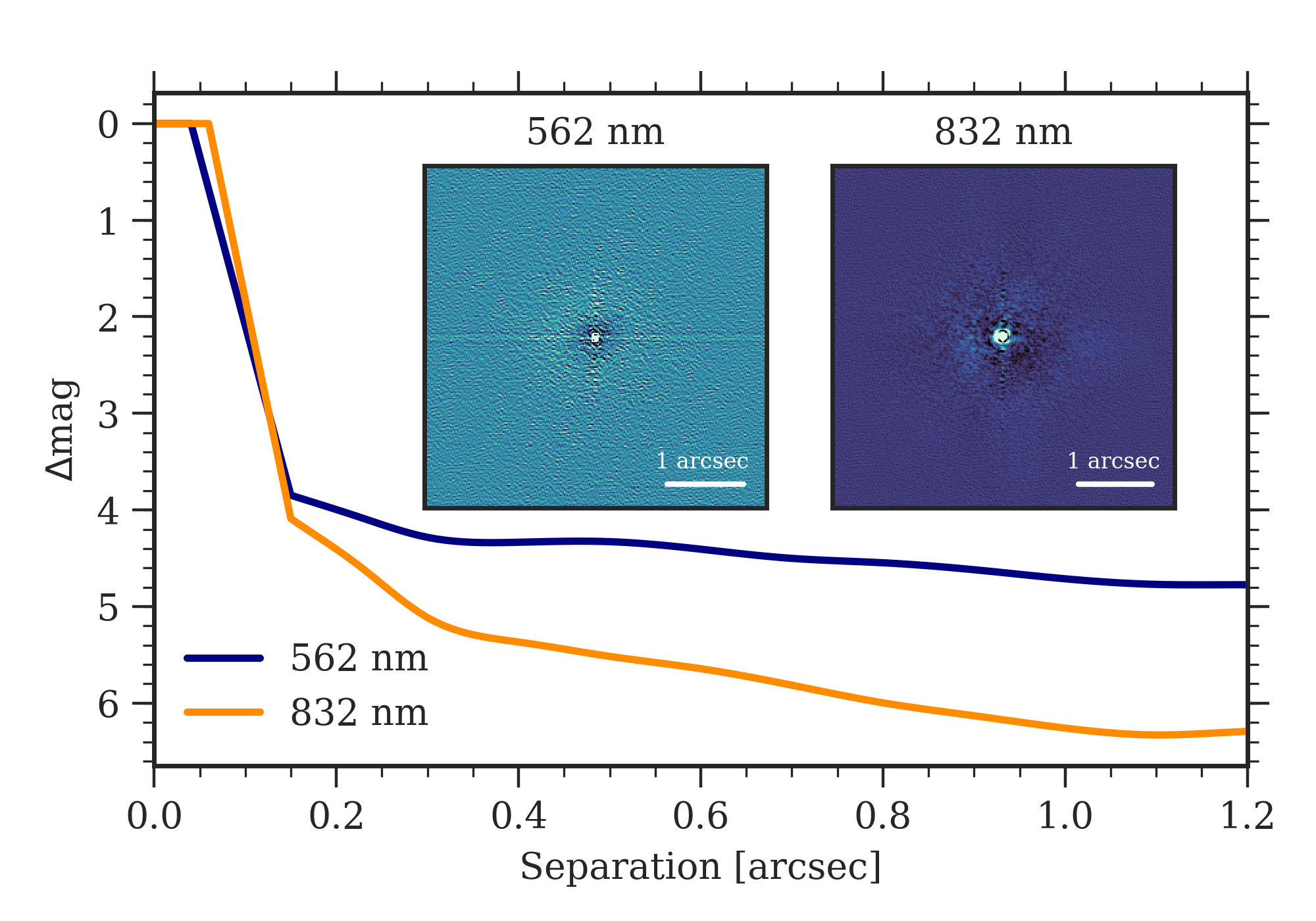} \\
\includegraphics[width=\hsize]{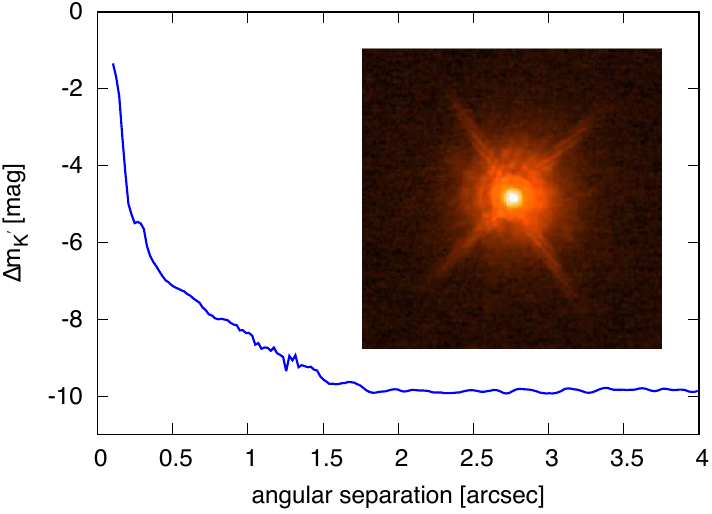}
\caption{{\it Top panel}: reconstructed images in the 562\,nm and 832\,nm narrow bands (inset) and their resulting $5\sigma$ contrast curves from WIYN/NESSI speckle interferometry. {\it Bottom panel}: Subaru/IRCS+AO188 combined saturated image and 5$\sigma$ contrast light curve of HD~119130.} 
\label{fig:imaging}
\end{figure}

We conducted speckle imaging observations of the host star using the NASA Exoplanet Star and Speckle Imager \citep[NESSI;][]{2016SPIE.9907E..2RS, 2018PASP..130e4502S}, mounted on the WIYN 3.5-m telescope at Kitt Peak Observatory. The observations were conducted simultaneously in two narrow bands centered at 562\,nm and 832\,nm. Following \citet{2011AJ....142...19H}, we collected and reduced the data, resulting in $4\farcs6 \times 4\farcs6$ reconstructed images of the host star. We did not detect any secondary sources in the reconstructed images, and we produced 5$\sigma$ background sensitivity limits from the reconstructed images using a series of concentric annuli (Fig.~\ref{fig:imaging}, top panel).

In addition, we observed HD~119130 using the Infrared Camera and Spectrograph \citep[IRCS;][]{2000SPIE.4008.1056K} and adaptive-optics system \citep[AO188,][]{2010SPIE.7736E..0NH} on the Subaru 8.2-m telescope. For HD~119130, two sequences were implemented with a five-point dithering using the $K^\prime$-band filter. The first sequence was for unsaturated frames for the absolute flux calibration, and we inserted a neutral-density filter with a transmittance $\sim$1\% to avoid saturation. We also obtained saturated frames to look for faint nearby companions as the second sequence without the neutral-density filter. The total integration times for unsaturated and saturated frames were 120\,s and 12\,s, respectively. 

We reduced all the IRCS frames in the standard manner as described in \citet{2016ApJ...820...41H}, and obtained median-combined images for the saturated and unsaturated frames, respectively. The combined unsaturated image suggested a full width at half maximum of $0\farcs078$, which is close to the diffraction limit. No nearby companion is seen in the saturated image. In order to estimate the detection limit of such companions, we computed the flux contrast based on the scatter of flux counts within the annulus centered at the target star. The bottom panel of Fig.~\ref{fig:imaging} shows a $5\sigma$ contrast curve as a function of angular separation from the central star, and its inset displays the target image with a field of view of $4\arcsec\times4\arcsec$.

\subsection{High-resolution spectroscopy}

We obtained a total of 24 measurements for HD~119130 from June 10th to July 18th 2018 with the CARMENES spectrograph \citep{CARMENES,CARMENES18}, installed at the 3.5-m telescope at the Calar Alto Observatory in Spain. The instrument consists of two channels: the visual channel (VIS) obtains spectra at a resolving power of $R = 94\,600$ in the wavelength range from $\SI{0.52}{\micro\metre}$ to $\SI{0.96}{\micro\metre}$, while the near-infrared (NIR) channel yields spectra of $R = 80\,400$ from $\SI{0.96}{\micro\metre}$ to $\SI{1.71}{\micro\metre}$. CARMENES performance, data reduction and wavelength calibration are described in \citet{Reiners18} and \citet{Kaminski18}. 

The dual-channel configuration of the CARMENES spectrograph is motivated by the desire to detect Earth-like planets around M dwarfs, whose redder spectral energy distribution requires red-optical and near-infrared coverage to derive precise RVs. However, due to the solar-like spectral type of HD~119130, we only use the VIS observations to measure RVs. We used SERVAL \citep{SERVAL}, a publicly available code based on least-squares fitting. SERVAL employs cross-correlation with a high signal-to-noise ratio template that is constructed by coadding all available spectra of the star, to derive RV values and several spectral indicators. In addition, we computed the cross-correlation function (CCF) using a weighted mask constructed from co-added CARMENES VIS spectra of HD~119130. We determined the RV, full width at half maximum, contrast, and bisector velocity span by fitting a Gaussian function to the final CCF, following the method described in \citet{Reiners18}. The final RVs are corrected for barycentric motion, secular acceleration, and nightly zero-points \citep[see][for details]{Luque18}.

\begin{figure}
\centering
\includegraphics[width=\hsize]{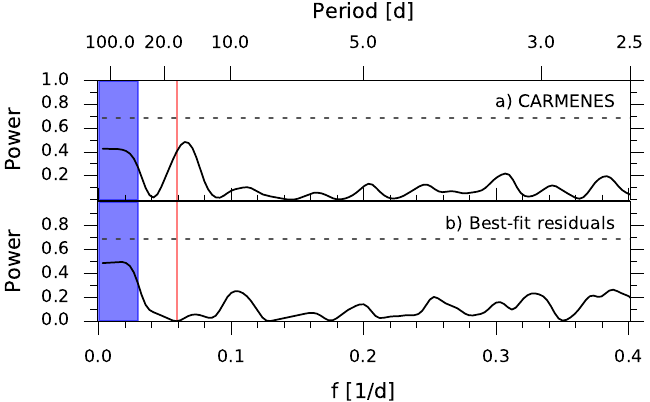}
\caption{{\it Top panel}: GLS periodogram of the CARMENES RVs. {\it Bottom panel}: GLS periodogram of the residuals from the best-fit model. The vertical red line marks the period of the planet derived from {\it K2} photometry. The dashed line corresponds to $\mathrm{FAP} = 10\%$ level. The shaded blue area indicates the period space longer than the time span of the observations.} 
\label{fig:gls}
\end{figure}

Due to the low declination of the star, HD~119130 was observed  from Calar Alto at relatively high airmasses (ranging from 1.5 to 1.9), which has a high impact on the telluric contamination of the spectra. Therefore, we do not consider for the RV computation those spectra whose signal-to-noise ratio per spectral sampling element, averaged over each order, is smaller than 40. Furthermore, to achieve the highest RV precision, we correct the spectra from telluric absorption using \texttt{Molecfit} \citep{Molecfit1,Molecfit2} following the method presented in \citet{Nortmann18} and \citet{Salz18}.

The CARMENES RV measurements are listed in Table~\ref{tab:RVs}. In total, 18 RVs were acquired using high signal-to-noise ratio telluric-free spectra, covering a timespan of 34\,d and with a median internal RV precision of $\sigma_{\mathrm{RV}} = 3.1\,\mathrm{m\,s^{-1}}$. By computing the generalized Lomb-Scargle periodogram \citep[GLS;][]{GLS} of the RVs (Fig.~\ref{fig:gls}a), we find a single peak at the expected frequency of the planet at $f = 0.06\,\mathrm{d^{-1}}$ ($P = 16.6\,\mathrm{d}$). The peak is broad due to the short temporal baseline of our measurements, but consistent within the frequency resolution with the planet candidate at $P \sim 17\,\mathrm{d}$. Although the peak has a false alarm probability (FAP) higher than 10\%, not enough to be claimed as a solid planet candidate from RVs alone, its consistency with the periodic signal detected in the {\it K2} photometry confirms the planetary nature of the RV variation. We also check for periodic signals at the expected planet frequency in the spectral indicators obtained with SERVAL and from the CCF, but find no evidence that the RVs periodicity is due to stellar effects. A peak is seen in H$\alpha$ and the CCF FWHM at $P \sim 3.5\,\mathrm{d}$  that may be related to activity. The GLS periodogram of the spectral indicators is shown in the Appendix (Fig.~\ref{fig:gls_all}) for completeness. We search also for possible correlations between the RVs and the other spectral indicators but find no evidence of Doppler shifts induced by line distortions. Hence, we conclude that the 17\,d-period transiting planet candidate from {\it K2} photometry is a bona fide planet also present in our CARMENES RV measurements.

\section{Stellar properties} \label{sec:star}

\begin{table}
\centering
\small
\caption{Stellar parameters of HD~119130.} \label{tab:star}
\begin{tabular}{lcr}
\hline\hline
\noalign{\smallskip}
Parameter 			    	& Value     		& Reference\tablefootmark{a} \\ 
\noalign{\smallskip}
\hline
\noalign{\smallskip}
\multicolumn{3}{c}{Coordinates and spectral type}\\
\noalign{\smallskip}
$\alpha$		    		& 13:41:30.30		& {\it Gaia} DR2 	\\
$\delta$		    		& -09:56:45.9		& {\it Gaia} DR2  	\\
SpT         				& G3~V              & {\citet{1999MSS...C05....0H}}		\\
$V$ [mag]                   & $9.917\pm0.013$   & \citet{2014AJ....148...81M}       \\
$J$ [mag]                   & $8.730\pm0.023$   & 2MASS       \\
\noalign{\smallskip}
\multicolumn{3}{c}{Parallax and kinematics}\\
\noalign{\smallskip}
$\pi$ [mas]		        	& $8.72 \pm 0.12$   	& {\it Gaia} DR2 		\\
$d$ [pc]		        	& $114.29^{+0.82}_{-0.81}$    	& \citet{2018AJ....156...58B} 		\\
$\mu_{\alpha}\cos\delta$ [$\mathrm{mas\,yr^{-1}}$]  & $-88.20 \pm 0.11$	& {\it Gaia} DR2  	\\
$\mu_{\delta}$ [$\mathrm{mas\,yr^{-1}}$]            & $1.67 \pm 0.12$	& {\it Gaia} DR2  	\\
$V_r$ [$\mathrm{km\,s^{-1}}]$             & $-28.18 \pm 0.16$	& {\it Gaia} DR2  	\\
\noalign{\smallskip}
\multicolumn{3}{c}{Photospheric parameters}\\
\noalign{\smallskip}
$T_{\mathrm{eff}}$ [K]		& $5725 \pm 65$         & This work         \\
                            & $5639^{+89}_{-98}$	& {\it Gaia} DR2 		\\
					        & $5645 \pm 57$	        & RAVE		\\
$\log g$				    & $4.30 \pm 0.15$	    & This work	    	\\
        				    & $4.33 \pm 0.04$	    & This work\tablefootmark{b} 	    	\\
					        & $3.90 \pm 0.08$	    & RAVE 		\\
{[Fe/H]}  				    & $0.07 \pm 0.05$	    & This work 		\\
					        & $0.00 \pm 0.09$   	& RAVE		\\
$v \sin i_\star$ [$\mathrm{km\,s^{-1}}$]	    & $4.6 \pm 1.0$	    & This work 		\\
\noalign{\smallskip}
\multicolumn{3}{c}{Physical parameters}\\
\noalign{\smallskip}
$M$ [M$_{\odot}$]			& $1.00 \pm 0.03$	        & This work		\\
                			& $1.08 \pm 0.12$	        & RAVE		\\
$R$ [R$_{\odot}$]		    & $1.09 \pm 0.03$	        & This work 		\\
                		    & $1.17^{+0.04}_{-0.04}$	& {\it Gaia} DR2 		\\
$\tau$ [Gyr]       	        & $6.8 \pm 2.3$	            & This work 		\\
\noalign{\smallskip}
\noalign{\smallskip}
\hline
\end{tabular}
\tablefoot{
\tablefoottext{a}{{\it References}. 2MASS: \citet{2MASS}; {\it Gaia} DR2: \citet{GaiaDR2}; RAVE: \citet{RAVE}.}
\tablefoottext{b}{Obtained from the mass and radius values derived with \texttt{PARAM 1.3}.}
}
\end{table}

\subsection{Photospheric parameters}\label{subsec:teff}

We used the co-added high-resolution CARMENES VIS spectrum in order to determine the physical parameters of the host star using the spectral analysis package \texttt{SME} v5.22 \citep{vp96,vf05}. \texttt{SME} is used to calculate, for a starting set of stellar parameters, a synthetic spectrum that is then fitted to the observed spectrum using a $\chi^2$ minimization procedure. \texttt{SME} makes use of a grid of recalculated stellar models, in our case the \texttt{ATLAS12} model atmospheres \citep{Kurucz2013}, which is a set of one-dimension models applicable to solar-like stars.

In order to determine the effective temperature, $T_{\mathrm{eff}}$, the profile of either of the strong Balmer line wings is fitted to the appropriate stellar spectrum models \citep{fuhrmann93,axer94,fuhrmann94,fuhr97b,fuhr97a}. This fitting procedure has to be carried out carefully since the determination of the level of the adjacent continuum can be difficult for modern high-resolution Echelle spectra where each order only contains a limited wavelength band \citep{fuhr97a}. The core of each Balmer line is excluded from the fitting process, since this part of the line profile originates above the photosphere and thus contributes at a different effective temperature.

We select parts of the observed spectrum that contain spectral features that are sensitive to the required parameters. We use the empirical calibration equations for Sun-like stars from \cite{Bruntt2010b} and \cite{Doyle2014} in order to determine the micro-turbulent ($v_{\rm mic} = 1.0 \pm 0.1\,\mathrm{km\,s^{-1}}$) and macro-turbulent ($v_{\rm mac} = 3.1 \pm 0.3\,\mathrm{km\,s^{-1}}$) velocities, respectively. The projected stellar rotational velocity, $v \sin i_\star$, and the metallicity, [Fe/H], are measured by fitting the profile of about 100 clean and unblended metal lines.

We find $T_{\rm eff} = 5725\pm65\,\mathrm{K}$, $\log g = 4.30\pm0.15$, $\mathrm{[Fe/H]} = 0.07\pm0.05$, and $v \sin i_\star = 4.6 \pm 1.0\,\mathrm{km\,s^{-1}}$. The effective temperatures derived from this work, as well as from {\it Gaia} DR2 \citep{GaiaDR2}, and RAVE \citep{RAVE} are within 1$\sigma$. We chose to use the value from our work since it is derived from high-resolution spectroscopy and for homogeneity, while the other values are derived from either low-resolution spectroscopy (RAVE) or photometry based on few wide bands ({\it Gaia}). Only the surface gravity does not agree within 1$\sigma$ with the results from \citet{RAVE}. All derived values and previous ones reported in the literature can be found in Table~\ref{tab:star}.

\subsection{Mass, radius and age of the host star}\label{subsec:mass}

To derive the physical parameters of HD~119130 we used \texttt{PARAM~1.3}\footnote{\url{http://stev.oapd.inaf.it/cgi-bin/param_1.3}}, a web interface for Bayesian estimation of stellar parameters using the \texttt{PARSEC} isochrones from \citet{PARSEC}. The required input is the effective temperature and metallicity of the star, together with its apparent visual magnitude and parallax. Following the method described in \citet{2008ApJ...687.1303G}, we found that the interstellar reddening is zero and did not correct the apparent visual magnitude for extinction. We use the spectroscopically derived photospheric parameters obtained in the previous section and the values reported in Table~\ref{tab:star}. For the {\it Gaia} parallax, we add quadratically 0.1\,mas to the nominal uncertainty to account for systematic uncertainties following \citet{Luri18}.

We derive a mass of $M = 1.00 \pm 0.03\,\mathrm{M_{\odot}}$, and a radius of $R = 1.09 \pm 0.03\,\mathrm{R_{\odot}}$, yielding a surface gravity of $\log g = 4.33 \pm 0.04$, in fairly good agreement with our spectroscopic value of $4.30 \pm 0.15$. The stellar models constrain the age of the star to be $6.8 \pm 2.3\,\mathrm{Gyr}$. We stress that the uncertainties on the derived parameters are internal to the stellar models used and do not include systematic uncertainties related to input physics. All derived values and previous ones reported in the literature can be found in Table~\ref{tab:star}.

\section{Joint analysis} \label{sec:fit}

\begin{table}  
\centering
\small
{\renewcommand{\arraystretch}{1.3}
\caption{Model comparison.}  \label{tab:models}
\begin{tabular}{llcccc}
  \hline
  \hline
  \noalign{\smallskip}
  
Model & Description & $K\,[\mathrm{m\,s^{-1}]}$ & $N_\mathrm{par}$ & $\chi^2_\nu$ & AIC     \\

  \noalign{\smallskip}
  \hline
  \noalign{\smallskip}

M0  & Keplerian orbit                  & $5.47_{-1.21}^{+1.02}$      & 11     & 1.56  & -2025.44  \\
M1  & M0  + lin. trend                 & $6.11_{-1.08}^{+1.06}$      & 12     & 1.19  & -2068.76  \\

    \noalign{\smallskip}
    \hline
    \end{tabular}}
\end{table}

To derive precisely the parameters of the HD~119130 system, we model simultaneously the photometric and spectroscopic data using the code \texttt{pyaneti} \citep{Pyaneti}. This software computes posterior distributions using Markov chain Monte Carlo methods based on Bayesian analysis. It uses the limb-darkened quadratic transit model by \citet{MandelAgol} to fit the {\it K2} light curves, and Keplerian orbits to model the RV data. We used for the joint fit the detrended {\it K2} light curve obtained with the method of \citet{Palle18} described in Section~\ref{sec:k2}. In order to account for the 30-min integration time of the photometric data, we resample the model using 10 iterations \citep[see][]{2010MNRAS.408.1758K}. The fitted parameters are the systemic velocity $\gamma_\mathrm{RV}$, the RV semi-amplitude $K$, the transit epoch $T_0$, the period $P$, scaled semi-major axis $a/R_\star$, planet-to-star radius ratio $R_p/R_\star$, impact parameter $b$, eccentricity $e$, longitude of periastron $\omega$, and the \citet{Kipping13} limb-darkening parametrization coefficients $q_1$ and $q_2$. We used wide-range non-informative uniform priors and the same likelihood as \citet{Barragan16}, except for the scaled semi-major axis, where we used Kepler's third law to set a Gaussian prior based on the stellar mass and radius derived in Section~\ref{subsec:mass}. To explore the parameter space, we created 500 independent chains for each parameter and checked their convergence using the Gelman-Rubin statistic. Once all chains converged, we ran 25\,000 more iterations with a thin factor of 50, leading to a posterior distribution of 250\,000 independent samples for each fitted parameter.

A simple joint fit to the data reveals that the residuals of the RVs exhibit a long-term component, as seen also in the GLS periodogram of Fig.~\ref{fig:gls}, where there is power at periods larger than the baseline of the observations. We considered two different scenarios regarding the nature of such long-term component: that it is physical and may be caused by a further companion and/or the rotation of the star, or that it is related to instrumental systematics. 

Considering the first scenario, the photometry does not show evidence of a second transiting companion during the observed window, or variability attributable to the rotational period of the star that may induce a long-term component in the RVs. However, it is possible that the hypothetical long-period companion does not transit or that the transit occurred before or after the {\it K2} observations. Alternatively, the long-term trend may be caused by systematic effects in the RVs, such as remaining effects from incomplete telluric decontamination. In either case, our data are not sufficient to infer the true nature of this long-term RV trend. 

In order to properly fit the current dataset, we tried different approaches to account for the RV long-term component described above. Thus, we performed the joint fit in two ways: i) disregarding any long-term trend (M0), and ii) including a linear trend term in the RV fit of the data (M1). Table~\ref{tab:models} shows the goodness of the fit for each model. The preferred model is M1, i.e., a Keplerian orbit with a linear trend term, with the lowest $\chi^2_\nu$ and Akaike information criterion \citep[AIC;][]{1974ITAC...19..716A}. We use the AIC in spite of the widespread Bayesian information criterion because it performs better in selecting the "true model" when the number of parameters is small \citep{10.1007}, as it is the case for our RV measurements. 

\begin{figure*}
\centering
\includegraphics[width=\hsize]{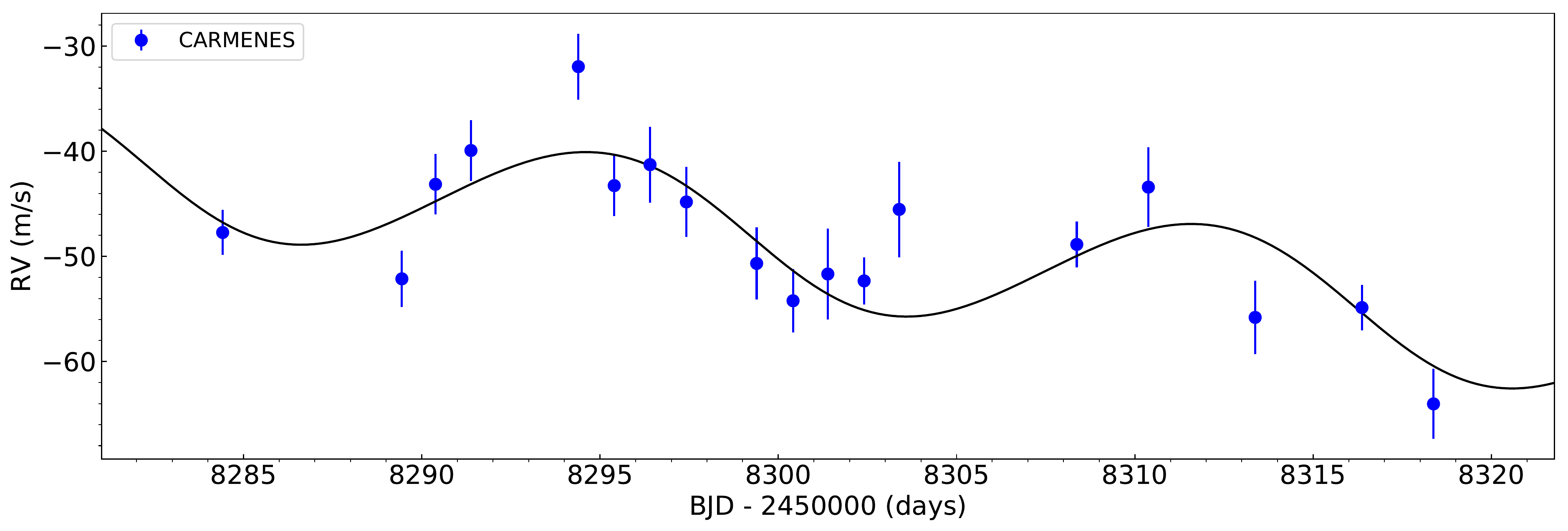} \\
\includegraphics[width=0.49\hsize]{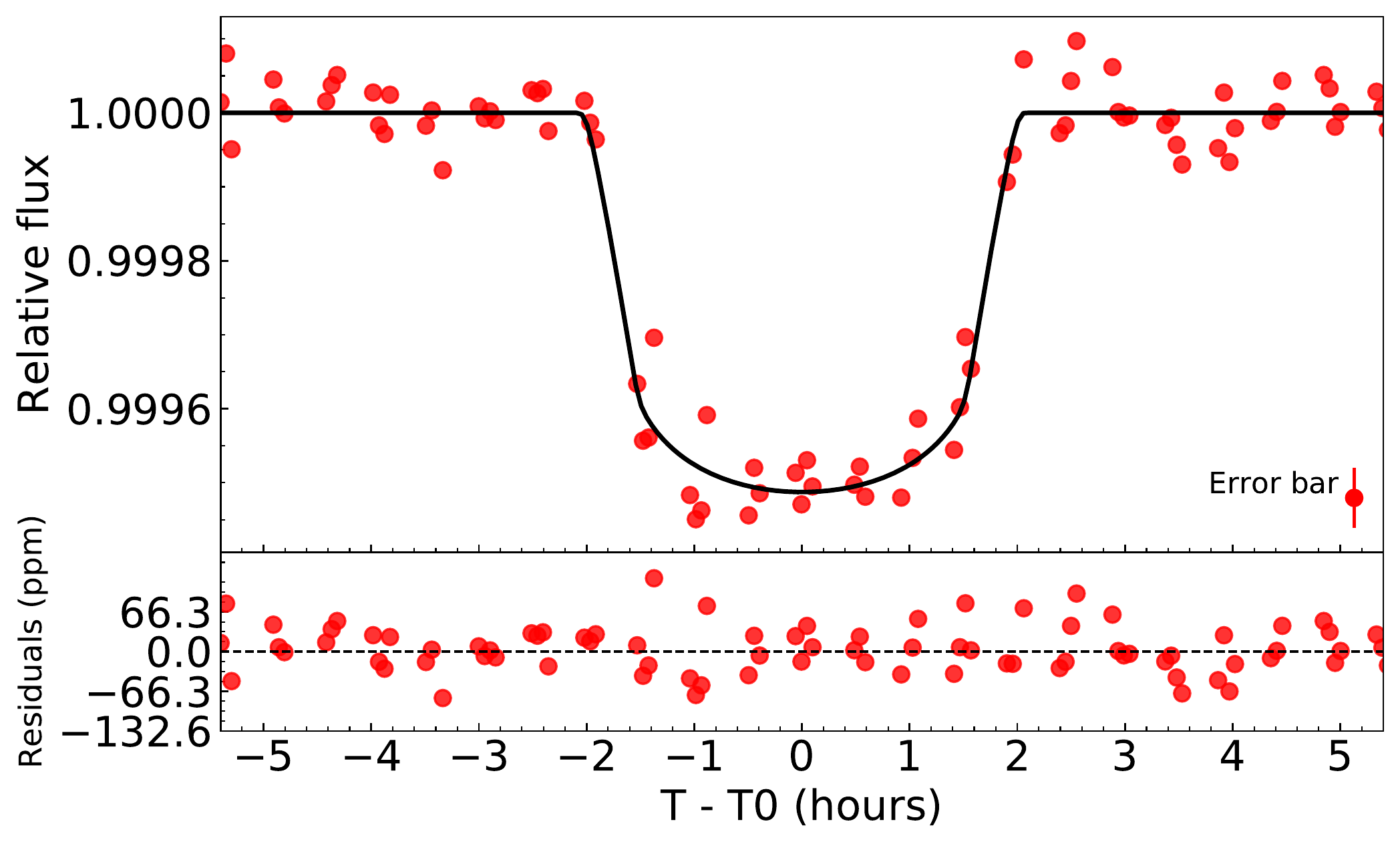}
\includegraphics[width=0.49\hsize]{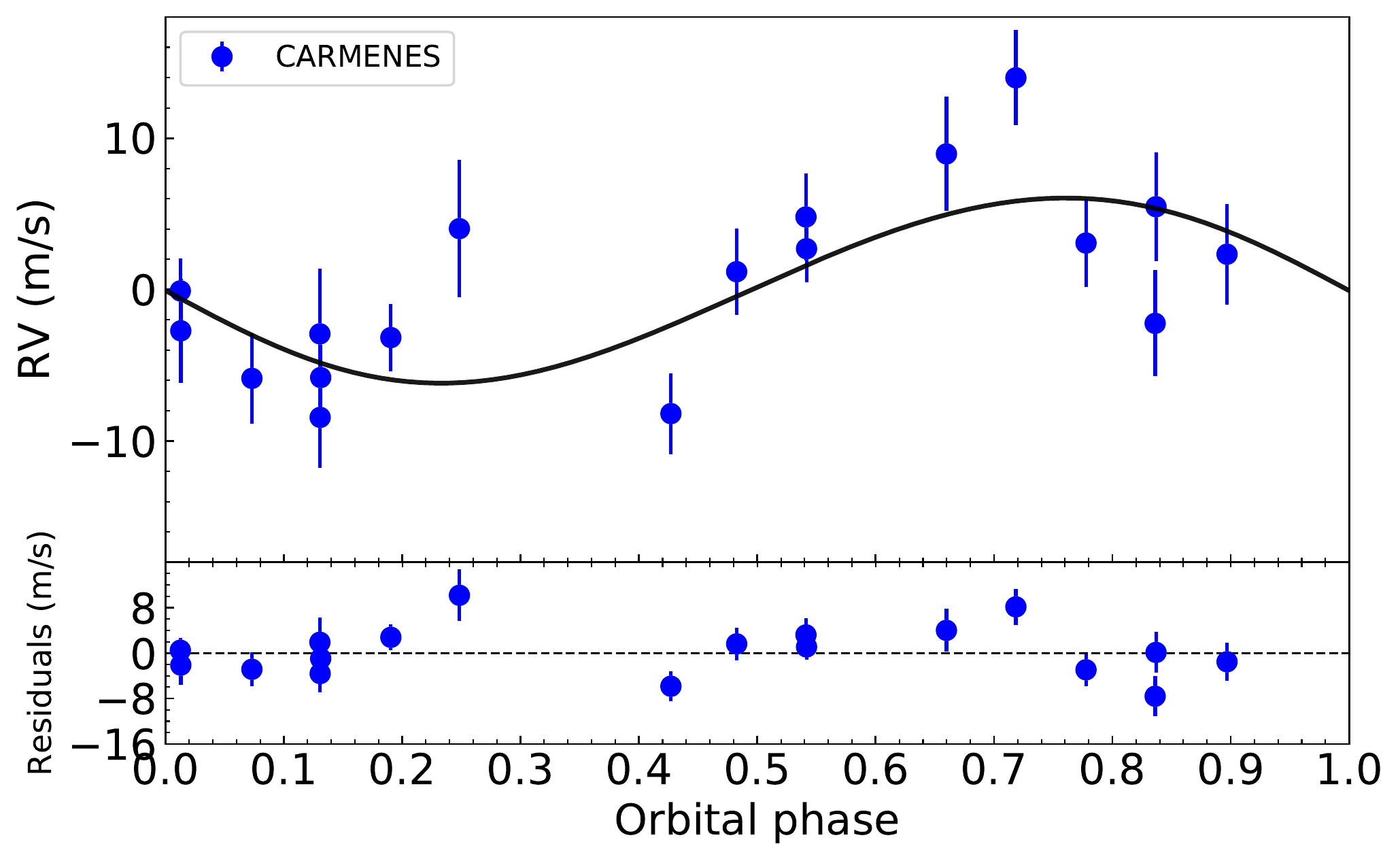}
\caption{{\it Top}: Time series of CARMENES RVs. {\it Bottom left}: Phase-folded {\it K2} light curve to the orbital period of HD~119130~b and residuals. {\it Bottom right}: Phase-folded RV curve to the same period and the linear trend subtracted. The solid black line in all panels indicate the preferred best-fit model M1.} 
\label{fig:rv_fit}
\end{figure*}

 The orbital parameters and their uncertainties from the photometric and spectroscopic joint fit of such model are listed in Table~\ref{tab:hd119130b}.  The large uncertainties in the limb-darkening coefficients arise from the shallow transit depth (550\,ppm) and the small number of data points and transits, especially during ingress and egress. Their posterior distributions are not symmetric and lean towards low values, but by fixing them interpolating the table of \citet{2011A&A...529A..75C} we retrieve the same values within $1\sigma$ uncertainties as those derived by \texttt{pyaneti} using uniform priors. Therefore, following \citet{2013A&A...549A...9C}, we choose to fit them in our joint analysis to achieve the best precision in the planet radius, impact parameter and scaled semi-major axis. Figure~\ref{fig:rv_fit} shows the RV time series along with the best fitting transit and RV phase-folded models.

\begin{table}  
\centering
{\renewcommand{\arraystretch}{1.3}
\caption{Orbital parameters of the preferred model M1.}  \label{tab:hd119130b}
\begin{tabular}{llc}
  \hline
  \hline
  \noalign{\smallskip}
  
Parameter       & Unit      & Value     \\

  \noalign{\smallskip}
  \hline
  \noalign{\smallskip}
  \multicolumn{3}{c}{Model parameters} \\
  \noalign{\smallskip}

$P$                             & d                             & \Pb[]         \\ 
$T_0$                           & BJD$-$2\,450\,000             & \Tzerob[]     \\  
$R_\mathrm{p}/R_{\rm \star}$    & \dots                         & \rrb[]        \\
$a/R_{\rm \star}$               & \dots                         & \arb[]        \\
$b$                             & \dots                         & \bb[]         \\
$e$                             & \dots                         & \eb[]     \\
$\omega$                        & deg                           & \wb[]     \\
$K$                             & $\mathrm{m\,s^{-1}}$          & \kb[]         \\
$\gamma_{\rm RV}$               & $\mathrm{m\,s^{-1}}$         & \CARMENES[]       \\
$\dot{\gamma}_{\rm RV}$         & $\mathrm{m\,s^{-1}\,d^{-1}}$  & \ltrend[]         \\
$q_1$                           & \dots                         &  \qone        \\
$q_2$                           & \dots                         & \qtwo         \\  

    \noalign{\smallskip}
    \hline
    \noalign{\smallskip}
    \multicolumn{3}{c}{Derived parameters} \\
    \noalign{\smallskip}
    
$M_\mathrm{p}$                  & $\mathrm{M_{\rm \oplus}}$     & \mpb[]        \\
$R_\mathrm{p}$                  & $\mathrm{R_{\rm \oplus}}$     & \rpb[]        \\
$\rho_{\rm p}$                  & $\mathrm{g\,cm^{-3}}$         & \denpb[]      \\
$g_{\rm p}$                     & $\mathrm{cm\,s^{-2}}$         & \grapparsb[]  \\
$\rho_{\star}$                  & $\mathrm{g\,cm^{-3}}$         & \denstrb[]        \\
$a$                             & au                            & \ab[]         \\
$i_o$                           & deg                           & \ib[]         \\
$\tau_{14}$                     & h                             & \ttotb[]      \\
$T_\mathrm{eq}$\tablefootmark{(a)} & K                             & \Teqb[]       \\
$F$                             & $\mathrm{F_{\oplus}}$         & \insolationb[] \\
$u_1$                           & \dots                         & \uone         \\
$u_2$                           & \dots                         & \utwo         \\

    \noalign{\smallskip}
    \hline
    \end{tabular}}

\tablefoot{
\tablefoottext{a}{Assuming Bond albedo equal zero.}
}
\end{table}

\section{Discussion and summary} \label{sec:discussion}

\begin{figure}
\centering
\includegraphics[width=\hsize]{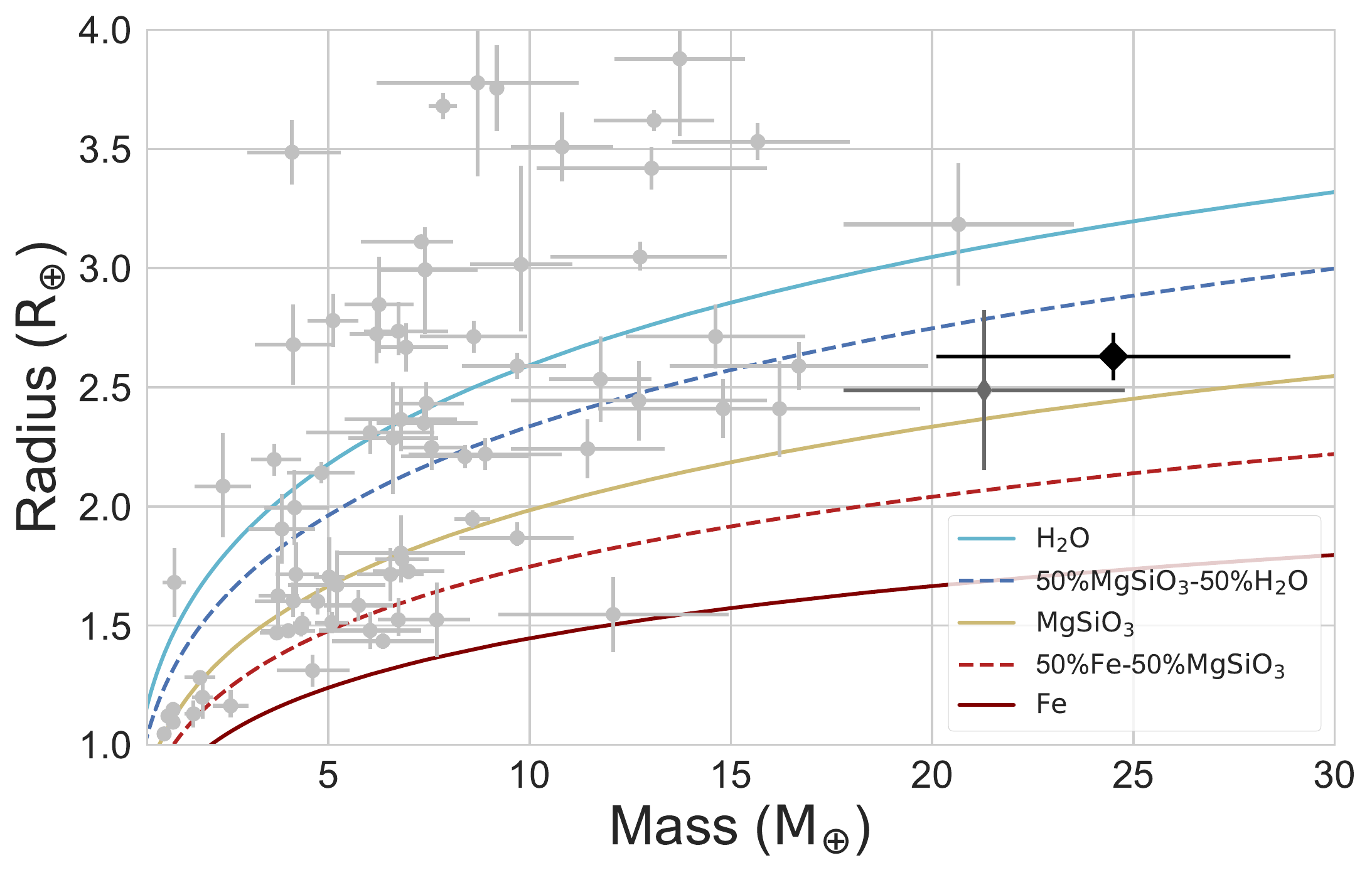}
\caption{Mass-radius diagram for all known planets with masses between 0.5--30\,$\mathrm{M}_\oplus$ and radius 1--4\,$\mathrm{R}_\oplus$ determined with a precision better than 30\%, comprising from Earth-like to sub-Neptune regimes. HD~119130~b is shown in black, while K2-66~b \citep{2017AJ....153..271S} is shown in grey. Data are taken from the TEPCat database of well-characterized planets \citep{2011MNRAS.417.2166S}. Theoretical models for the planet's internal composition are taken from \citet{2016ApJ...819..127Z}. } 
\label{fig:massradius}
\end{figure}

We determine that HD~119130~b has a mass of $M_\mathrm{p} = 24.5_{-4.4}^{+4.4}\,\mathrm{M_\oplus}$ and a radius of $R_\mathrm{p} = 2.63_{-0.10}^{+0.11}\,\mathrm{R_\oplus}$, which corresponds to a bulk density $\rho_\mathrm{p} = 7.4_{-1.5}^{+1.6}\,\mathrm{g\,cm^{-3}}$. Figure~\ref{fig:massradius} shows the masses and radii of all confirmed planets whose precision in both parameter determinations is better than 30\%. With such physical parameters, HD~119130~b becomes one of the densest sub-Neptune planet known to date, joining a rare population with only K2-66~b \citep{2017AJ....153..271S} as a comparable case. For these planets, mass-radius relations such as the one presented by \citet{2014ApJ...783L...6W} do not apply. Instead, we see the remarkable differences in composition of planets with masses ranging from 10 to 30\,M$_\oplus$. 

HD~119130~b is consistent with a pure silicate composition using \citet{2016ApJ...819..127Z} models. More complex three-layer models provide very similar results \citep{2013PASP..125..227Z}. Also, following \citet{2014ApJ...792....1L}, we determine that if the planet has an H/He envelope atop the rocky-composition core, the former would represent less than 1\% of the planet's mass, which is in agreement with the statistical study of \citet{2015ApJ...806..183W} using sub-Neptune planet candidates from {\it Kepler}. In the same study, the authors also stress the lack of a deterministic mass-radius relationship and the necessity of reliable mass measurements for the sub-Neptune planet population. 

The only other known ultra-dense sub-Neptune K2-66~b orbits very close to its host star and receives a high level of irradiation, placing it in the photoevaporation desert defined by \citet{2016NatCo...711201L}. As a consequence, together with the fact that K2-66 is evolving up the subgiant branch, \citet{2017AJ....153..271S} concluded that photoevaporation stripped off the low-density volatiles in the planet envelope leading to the high density measured today. We investigate if the high density of HD~119130~b could also be due to the absence of a substantial atmosphere caused by evaporation, as it would be expected considering its location above the radius gap \citep{Fulton18,2018MNRAS.479.4786V}. Following the formulation in \citet{2007A&A...461.1185L}, we derive a mass loss rate from extreme ultraviolet radiation of the star of only $0.005\,\mathrm{M_\oplus}$ over the age of the star, which is insufficient to reduce significantly the radius of the planet. In addition, due to its period of 17\,d and an incident flux of $67\,\mathrm{F_\oplus}$, HD~119130~b is far from the photoevaporation desert.

Consequently, HD~119130~b likely formed with high density, similar to K2-110~b \citep[$\rho_\mathrm{p}=5.2\pm1.2\,\mathrm{g\,cm^{-3}}$,][]{2017A&A...604A..19O}. The remaining question is then how such a dense planet located at $0.13\,\mathrm{au}$ from its host star may have formed. Formation in situ can be ruled out, as a disk mass enhancement by a factor of $\sim 40$ above the minimum-mass solar nebula is needed to form HD~119130~b \citep{2014ApJ...795L..15S}. However, no significant enhancement is needed if the planet formed at larger distances ($> 2\,\mathrm{au}$), making inward migration a plausible explanation. There are two main mechanisms to trigger inward migration: via interactions with the gaseous disk or dynamical interactions with another body. The latter type arises from the gravitational force exerted by a sufficiently massive body, which can also be expressed in terms of a torque. This torque alters the angular momentum of the planet's orbit, resulting in a variation of the orbital elements, particularly, a decrease over time of the semi-major axis. Particularly, the migration could indeed be caused by Kozai-Lidov oscillations \citep{2014Sci...346..212D,2017MNRAS.468.3000M}; these could be excited by the possible long-period companion suggested by a linear trend in the RV data. The derived slope, in case it is of planetary nature, suggests a long-period companion with a minimum mass on the order of $33\,\mathrm{M_\oplus}$ \citep[see Equation~1 in][]{2015ApJ...800...22F}. 

Multiplanetary systems sculpted by Kozai-Lidov mechanisms are expected to exhibit significant mutual inclinations and this could be the reason that the further companion is not detected in the {\it K2} light curve. The eccentricity of the transiting planet cannot be well constrained by the current RV data, but it is reasonable to assume that the eccentricity is small but non zero, given that there is only one single transiting planet. \citet{vanEylen19} found that there is an eccentricity spread for this kind of systems, and that they are well described by the positive half of a Gaussian, peaking at zero and with a width of $0.32 \pm 0.06$. However, to prove the proposed scenario, more measurements would be needed to constrain the eccentricity of the planet and reveal the nature of the slope in the RV time series.

\begin{acknowledgements}

This paper includes data collected by the {\it K2} mission. Funding for the {\it K2} mission is provided by the NASA Science Mission directorate.

This work has made use of data from the European Space Agency (ESA) mission {\it Gaia} (\url{https://www.cosmos.esa.int/gaia}), processed by the {\it Gaia} Data Processing and Analysis Consortium (DPAC, \url{https://www.cosmos.esa.int/web/gaia/dpac/consortium}). Funding for the DPAC has been provided by national institutions, in particular the institutions participating in the {\it Gaia} Multilateral Agreement.

CARMENES is an instrument for the Centro Astron\'omico Hispano-Alem\'an de Calar Alto (CAHA, Almer\'ia, Spain) funded by the German Max-Planck-Gesellschaft (MPG), the Spanish Consejo Superior de Investigaciones Cient\'ificas (CSIC), the European Union through FEDER/ERF~FICTS-2011-02 funds, and the members of the CARMENES Consortium.

R.L. has received funding from the European Union’s Horizon 2020 research and innovation program under the Marie Skłodowska-Curie grant agreement No.~713673 and financial support through the “la Caixa” INPhINIT Fellowship Grant for Doctoral studies at Spanish Research Centres of Excellence, “la Caixa” Banking Foundation, Barcelona, Spain. 
This work is partly financed by the Spanish Ministry of Economics and Competitiveness through grants ESP2013-48391-C4-2-R and AYA2016-79425-C3, and supported by the Japan Society for Promotion of Science (JSPS) KAKENHI Grant Number JP16K17660. Funding for the Stellar Astrophysics Centre is provided by The Danish National Research Foundation (Grant agreement no.: DNRF106).
K.W.F.L acknowledges the support of the DFG priority program SPP~1992 "Exploring the Diversity of Exoplanets in the Mass-Density Diagram" (RA~714/14-1). M.F. and C.M.P. gratefully acknowledge the support of the 
Swedish National Space Agency.
\end{acknowledgements}

\bibliographystyle{aa} 
\bibliography{biblio} 


\begin{appendix} 

\section{Frequency analysis of spectral indicators.} 

\begin{figure}
\centering
\includegraphics[width=\hsize]{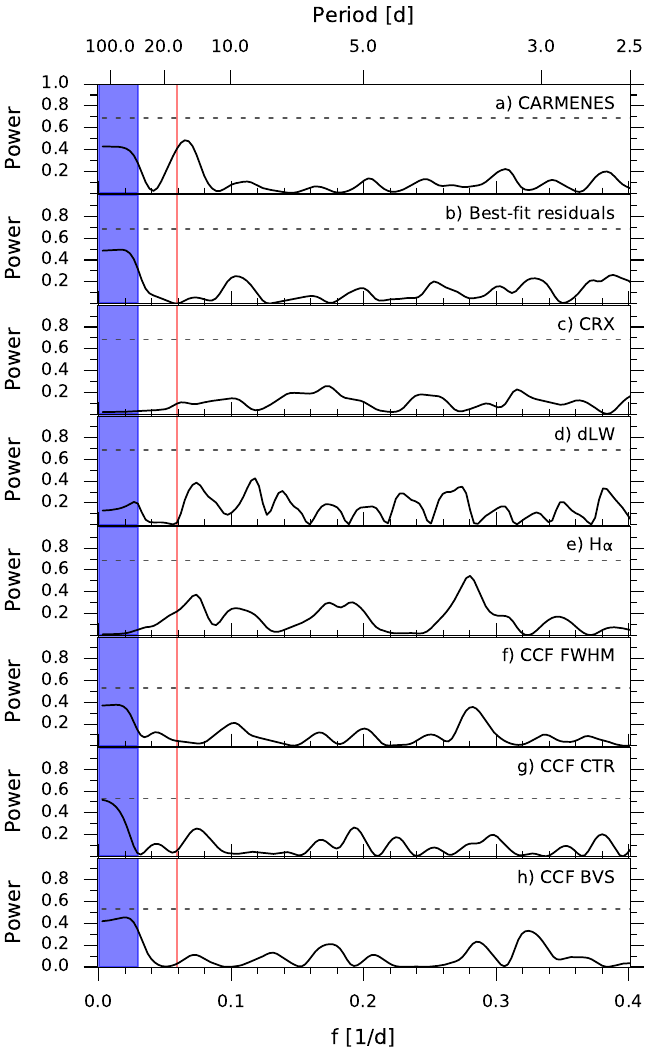}
\caption{GLS periodograms of CARMENES RVs (a) and its residuals (b) after removing the planet signal at $f = 0.06\,\mathrm{d^{-1}}$ derived from {\it K2} photometry, marked in red. Panels (c--e) show periodograms of the chromatic index, differential line width, and H$\alpha$ index, while panels (f--h) show periodograms for the cross-correlation function full-width half-maximum, contrast, and bisector velocity span. The horizontal dashed line shows the theoretical FAP level of 10\%. The shaded blue area indicates the period range longer than the timespan of the observations.}  \label{fig:gls_all}
\end{figure}

\section{Radial velocity measurements}

\begin{table}
\centering
\caption{Radial velocities and formal uncertainties of HD~119130}\label{tab:RVs}
{\renewcommand{\arraystretch}{1.0}
 \renewcommand{\tabcolsep}{0.03\textwidth}
\begin{tabular}{ccc}
\hline\hline
\noalign{\smallskip}
$\mathrm{BJD}$ & RV & $\mathrm{\sigma_{RV}}$ \\
  & [$\mathrm{m\,s^{-1}}$] & [$\mathrm{m\,s^{-1}}$] \\
\noalign{\smallskip}
\hline
\noalign{\smallskip}
2458284.416 &   0.86 & 2.14 \\
2458289.442 &  -3.55 & 2.67 \\
2458290.385 &   5.44 & 2.87 \\
2458291.379 &   8.66 & 2.89 \\
2458294.392 &  16.63 & 3.14 \\
2458295.398 &   5.32 & 2.89 \\
2458296.403 &   7.31 & 3.61 \\
2458297.422 &   3.76 & 3.32 \\
2458299.393 &  -2.09 & 3.43 \\
2458300.413 &  -5.64 & 3.00 \\
2458301.389 &  -3.09 & 4.31 \\
2458302.407 &  -3.75 & 2.23 \\
2458303.391 &   3.05 & 4.54 \\
2458308.372 &  -0.28 & 2.20 \\
2458310.379 &   5.17 & 3.78 \\
2458313.374 &  -7.22 & 3.50 \\
2458316.370 &  -6.29 & 2.15 \\
2458318.377 & -15.45 & 3.33 \\
\noalign{\smallskip}
\hline
\end{tabular}
}
\end{table}

\end{appendix}

\end{document}